\documentclass[5p]{elsarticle}
\usepackage{color} 

\usepackage{hyperref}

\graphicspath{{./images/}}

\begin{document}

\begin{frontmatter}

\title{Effect of long- and short-range interactions on the thermodynamics of dipolar spin ice}

\author[fefu]{Yuriy Shevchenko\corref{mycorrespondingauthor}}
\cortext[mycorrespondingauthor]{Corresponding author}
\ead{shevchenko.ya@dvfu.ru}

\author[fefu]{Aleksandr Makarov}
\ead{makarov.ag@dvfu.ru}

\author[fefu,ras]{Konstantin Nefedev}
\ead{nefedev.kv@dvfu.ru}

\address[fefu]{School of Natural Sciences, Far Eastern Federal University, Vladivostok, Russian Federation}
\address[ras]{Institute of Applied Mathematics of Far Eastern Branch, Russian Academy of Science, 7 Radio Str, Vladivostok, Russian Federation}

\begin{abstract}
The thermodynamic properties of dipolar spin ice on square, honeycomb and shakti lattices in the long-range and short-range dipole interaction models are studied. 
Exact solutions for the density of states, temperature dependencies of heat capacity, and entropy are obtained for these lattices with a finite number of point dipoles by means of complete enumeration. 
The magnetic susceptibility and average size of the largest low-energy cluster are calculated for square spin ice by means of Wang-Landau and Metropolis methods. 
We show that the long-range interaction leads to a blurring of the energy spectrum for all considered lattices. The inclusion of the long-range interaction leads to a significant change in the thermodynamic behaviour. 
An additional peak of heat capacity appears in the case of the honeycomb lattice. 
The critical temperature shifts in the direction of low or high temperatures; the direction depends on the lattice geometry. 
The critical temperature of the phase transition of square spin ice in the long-range model with frustrated ground states is obtained with the Wang-Landau and Metropolis methods independently.
\end{abstract}

\begin{keyword}
Artificial dipolar spin ice\sep
Honeycomb\sep
Shakti\sep
Density of states\sep
Entropy\sep
Heat capacity\sep
Magnetic susceptibility\sep
Order parameter\sep
Complete enumeration\sep 
Wang-Landau\sep 
Metropolis\sep
\end{keyword}

\end{frontmatter}


\section{Introduction}

In the middle of last century, Anderson revealed the notable likeness of spinel structure with that of water ice \cite{anderson1956ordering}, and later the similarity with the pyrochlore lattice was discovered \cite{PhysRevLett.79.2554}. As a consequence, such Ising antiferromagnet spin systems were named ''spin ice''. Three-dimensional crystals of spin ice have a whole set of interesting phenomena depending on temperature and an external magnetic field, which are usually explained by the presence of frustrations \cite{PhysRevLett.79.2554, Snyder2001, PhysRevB.68.180401, Ramirez1999, PhysRevB.78.174416}. 
However, the detailed control of the thermodynamic states of 3D spin ice, aimed at researching frustration phenomena and detailed consideration of configurations (conformations), encounters difficulties if well-known experimental methods are used. 
Today, spin ice prototypes are actively investigated. 
The artificial spin ice (ASI) system consists of an ordered or quasi-ordered array of ferromagnetic single domain nanoparticles with significant shape anisotropy. ASI is usually used as a 2D representation of spin ice.

ASI single-domain ferromagnetic nanoislands have magnetic moments that behave like Ising superspins \cite {wang2006artificial}. 
They are interesting both from the experimental and theoretical points of view. 
The interest in the preparation and study of such systems is due to the practically unlimited possibilities of array geometry. 
Modern nanolithography techniques allow one to create nanoparticle arrays of microscopic size. 
The monitoring of thermodynamic states and concrete magnetic configurations has become possible due to advances in magnetic force microscopy (MFM) and other highly sensitive techniques; for example, photo emission electron microscopy (PEEM) and magnetic circular X-ray dichroism (XMCD).
Initially, ASI was designed to simulate the behaviour of pyrochlore samples, but today it has become a fully independent platform for checking the well-known models of statistical mechanics, such as the Ising model \cite {unnar_artificial_ising}, and even those that do not have analogues in natural materials \cite {triangular_asi}.
There are possibilities to vary the lattice parameter, array geometry, magnetic moments of nanoparticles, and geometry of nanoislands during sample preparation to bring the energy barriers between individual microstates to the $k_ {B} T$ order or greater \cite{PhysRevLett.111.057204}. 
In other words, it is possible to ''freeze'' the magnetic states, i.e. to make them more static at the fixed $T$ or vice versa, to reduce the  ''lifetime'' of observable microstates.

Artificial square spin ice (ASSI) was one of the first objects that could be related to the class of frustrated systems \cite{wang2006artificial}. 
It remains attractive because the presence of frustrations does not forbid the existence of a phase transition. 
Actually, the geometry of ordering of magnetic moments at low temperatures has been well studied, but unfortunately the majority investigations of ASSI models were made in approximation of nearest neighbours.
Therefore, ASSI can be used as a test system to verify the applicability of models with a short range of interaction. 
In addition, here, we use ASSI to present our approach to the description of states using the cluster parameter of order.

Today, new geometries of lattice of ASI are known. Some of them have been created and studied experimentally; for example, honeycomb \cite{PhysRevLett.106.207202,PhysRevB.80.140409} and others \cite{morrison2013unhappy,chern2013degeneracy,gilbert2014emergent,gilbert2015emergent,morrison2013unhappy}.
In our opinion, first small samples of ASI must be studied, which can be produced experimentally and studied theoretically to understand the origin of ''new physics''.

\begin{figure}

\begin{minipage}[h]{1\linewidth}
\centering{\includegraphics[width=1\linewidth]{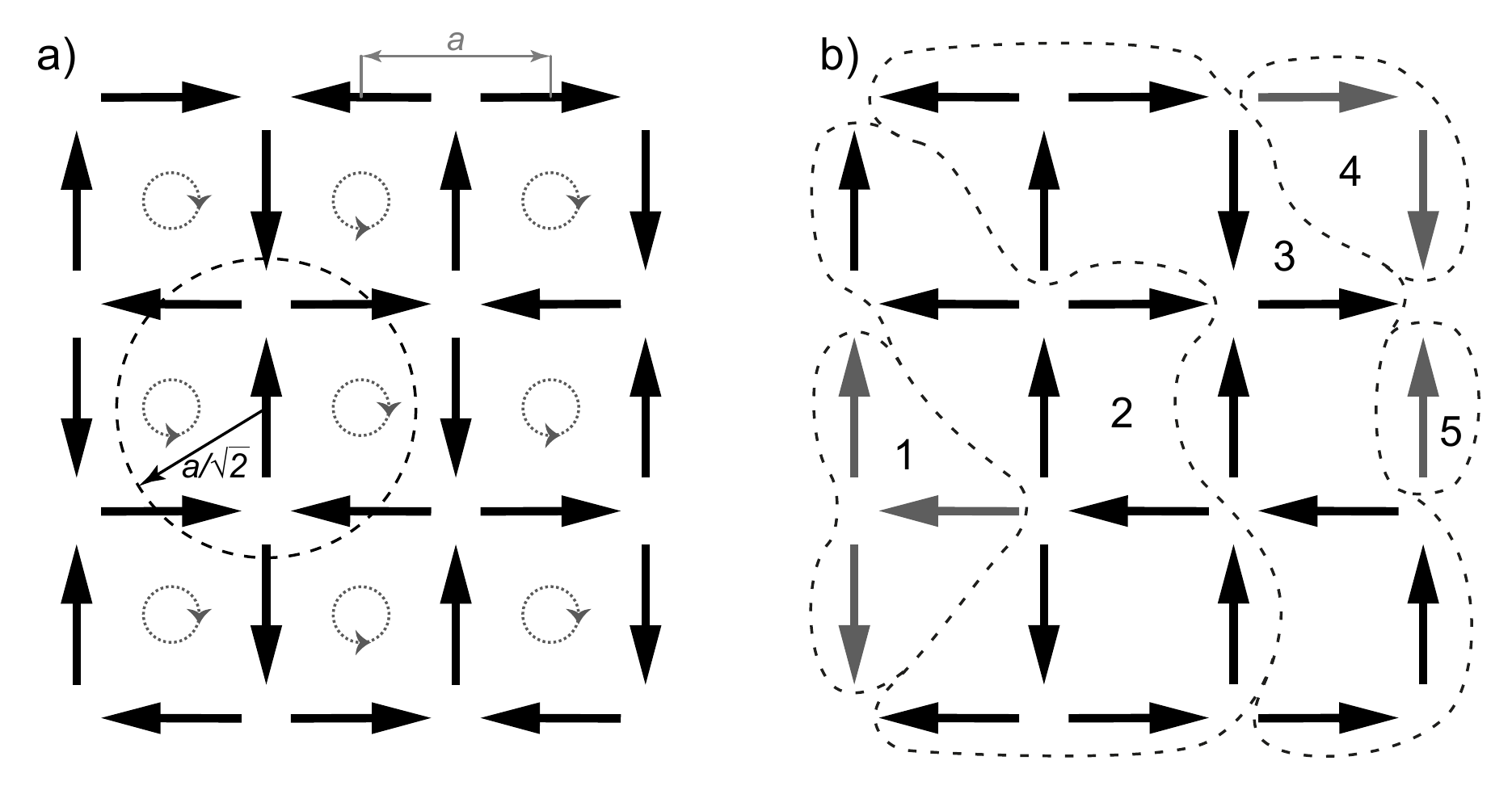}}
\end{minipage}
\vfill
\begin{minipage}[h]{1\linewidth}
\centering{\includegraphics[width=1\linewidth]{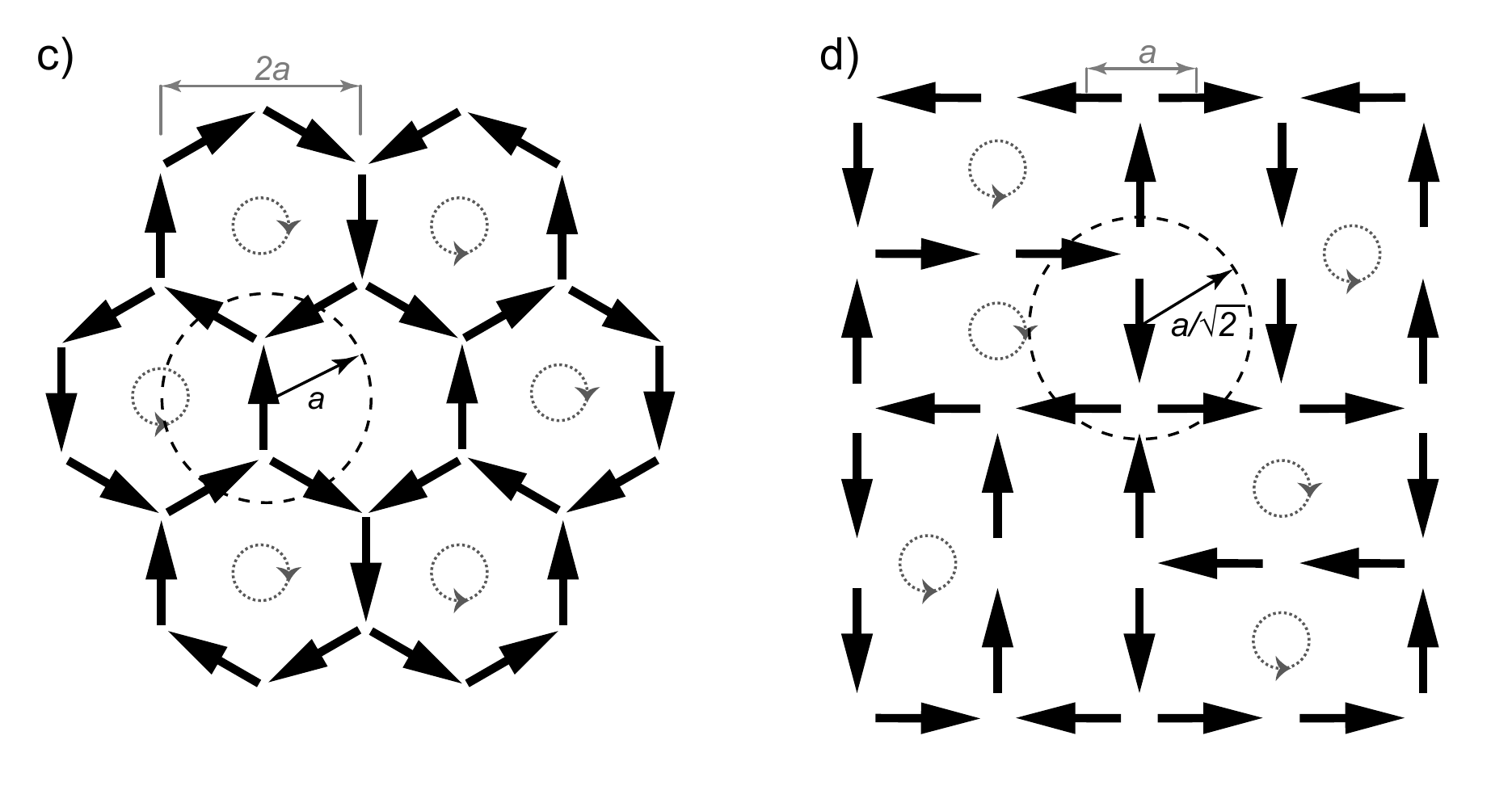}}
\end{minipage}

\caption{Examples of dipole lattices in ground states. a) One of two equiprobable ground states of ASSI. b) Proposed way to clusterise ASSI. c) Honeycomb-ASI ground state. d) Shakti-ASI ground state.}
\label{ASI_example}

\end{figure}

In our model, an ASI consists of magnetic moments placed in the ribs of a lattice (Figure \ref{ASI_example}).
As a result of dipolar interactions, the magnetic moments of neighbouring dipoles prefer the ''head-to-tail'' ordering.
The configurations of dipoles in local energy minimum satisfy the ''ice rule'' \cite {PhysRevLett.79.2554} Figure \ref{ASI_example}a (two inside, two outside) in each vertex for ASSI, and ''quasi ice rule'' (two in, one out, or two out, one in) ffor shakti-ASI and honeyocomb-ASI lattices, as shown in Figures \ref{ASI_example}c and \ref{ASI_example}d.

ASI is usually made of thin film materials, for instance permalloy ($Ni_{80\%}Fe_{20\%}$) or pure cobalt. 
The total volume of a nanoisland is $\sim 10^5$ $nm^3$ or less to be strictly the single domain. 
The magnetisation is $\sim 10^{7} \mu_B$. 
Elongation of the island produces the shape anisotropy and the parallel ordering of the atom's spins along the longest side. The shape anisotropy is the main reason behind the reduction of superspin vector components. 
The Ising-like behaviour of the island's magnetic moment is confirmed by MFM-images \cite{RevModPhys.85.1473}, and independent research into the nanoparticle’s magnetic states by means of PEEM-XMCD methods \cite{PhysRevLett.111.057204, Kapaklis.Vassilios2014}.
Due to the nanoscale employed, it would be logical to consider a nanoisland moment as a magnetic point dipole, and the interaction among themselves as dipole-dipole.

The influence of interaction length on the residual entropy and thermodynamic properties of a many-body system is an old problem, which has been studied for a long time in different contexts. 
In 1D and 2D systems, the long-range order of electrical or magnetic moments in lattice geometry is possible \cite{Heyderman2013, MALOZOVSKY1991127}. 
As will be shown below, the set of ground state spin configurations of dipoles on a lattice has a strong dependence on the geometry of the nanoparticle array and the radius of interaction. 
The long-range interaction strength decays as $r^{-3}$, and is sometime cut to the nearest neighbours \cite{levis2013thermal}, or to the next-nearest neighbours \cite{1367-2630-14-1-015008}. 
Differences in the thermodynamic properties of short-range and long-range models are possible: see for example \cite{PhysRevLett.106.057209}. 
The long-range dipole interactions, tails beyond the nearest neighbours, are important \cite{Mengotti}.
The most probable magnetic configurations, temperature behaviour and response to external influences can all strongly depend on the approximation level and error \cite{PhysRevB.73.020405,de2014magnetic,gliga2015broken}. 
The question ''could dipole systems with long interaction be studied in the frame of a short-range model?'' is actively discussed in the literature. 
However, in some cases, short-range models work. 
In \cite{Bramwell2006, PhysRevB.63.184412, PhysRevLett.95.217201} the unexpectedly good nearest neighbours approach allows the properties of spin ice to be described. 

One of the most effective theoretical methods of thermodynamics research in this area is with the Monte-Carlo (MC) method. 
It is usually supposed that the Metropolis algorithm with single-spin flip dynamics is insufficient at low temperatures due to critical slowdown; however, there are some approaches that allow this barrier to be overcome \cite{PhysRevLett.87.067203, PhysRevB.90.220406}. 
It is interesting to study the thermodynamics of ASI with long- and short-range dipole interaction. 
The aim of this work is to research the effect of interaction length on the thermodynamics of ASI on honeycomb, shakti and square lattices.

\section{Dipolar model}

As stated before, artificial spin ice is an array of single-domain ferromagnetic nanoislands. 
In our ASI model, we idealise magnetic dipole moment as a macrospin or an Ising superspin, due to the strong shape anisotropy (hereafter, for simplicity, we will use ''spin'' instead of ''superspin'' or ''macrospin''). 
Of course, the ideal dipole is an approximation and simplification.
The real nanoisland can have edge magnetic phenomena, whose intensity depends on the lattice topology and proximity of nanoislands in the vertex. 
It must be noted that, in the dipolar model, the ''magnetic charges'' are totally concentrated at the centres of the lattice ribs; therefore our models have a significant difference from the vertex models. 
XMCD images \cite{PhysRevB.78.144402}, allow us to conclude that the density of the magnetisation is uniform over the entire length of the nanoisland. This confirms the single-domain nature of the islands and demonstrates the dominance of shape anisotropy \cite{wang2006artificial}.
The single-domain state and orientation of magnetic moment directed along the long axis lead to effective Ising-like behaviour. 
Also, we should note that our model (ideal dipole) is also a simplification. 
It does not include relaxation effects at the tips of the particles. 
In our model, the nanoisland is on the rib of the lattice, and its longest side is significantly less than the length of rib.

\cite{Nefedev2010260, Nefedev2011325, Ivanov2012222,Ivanov2011} present the results of numerical simulations of magneto-force microscopy experiments, and the results of studying the influence of shape anisotropy on the behaviour of the nanoislands. 
It is possible to make the conclusion that the MFM images show the absence of multi-domain states, and it is also possible to ignore the declination from uniform magnetisation of the nanoislands \cite{RevModPhys.85.1473}. 
All our solutions work are obtained in the proposition of equilibrium, i.e. under the condition that all energetic barriers, which are controlled by shape anisotropy or other types of magnetic anisotropies, have been overcome.

The energy of the dipole interaction for the $ij$ spin pair is:
\begin{equation}
E^{ij}_{dip}=D \left(
\frac{
	(\vec{m}_i\vec{m}_j)
}{
	\vert\vec{r}_{ij}\vert^3
}
-3
\frac{
	(\vec{m}_i\vec{r}_{ij})
	(\vec{m}_j\vec{r}_{ij})
}{
	\vert\vec{r}_{ij}\vert^5
}\right),
\end{equation}
where 
$D=\mu^2/a^3$ - dimensional coefficient,
$\mu$ - total magnetic moment of the island, 
$a$ - lattice parameter.

\begin{figure}

\vfill
\begin{minipage}[h]{0.49\linewidth}
\center{\includegraphics[width=1\linewidth]{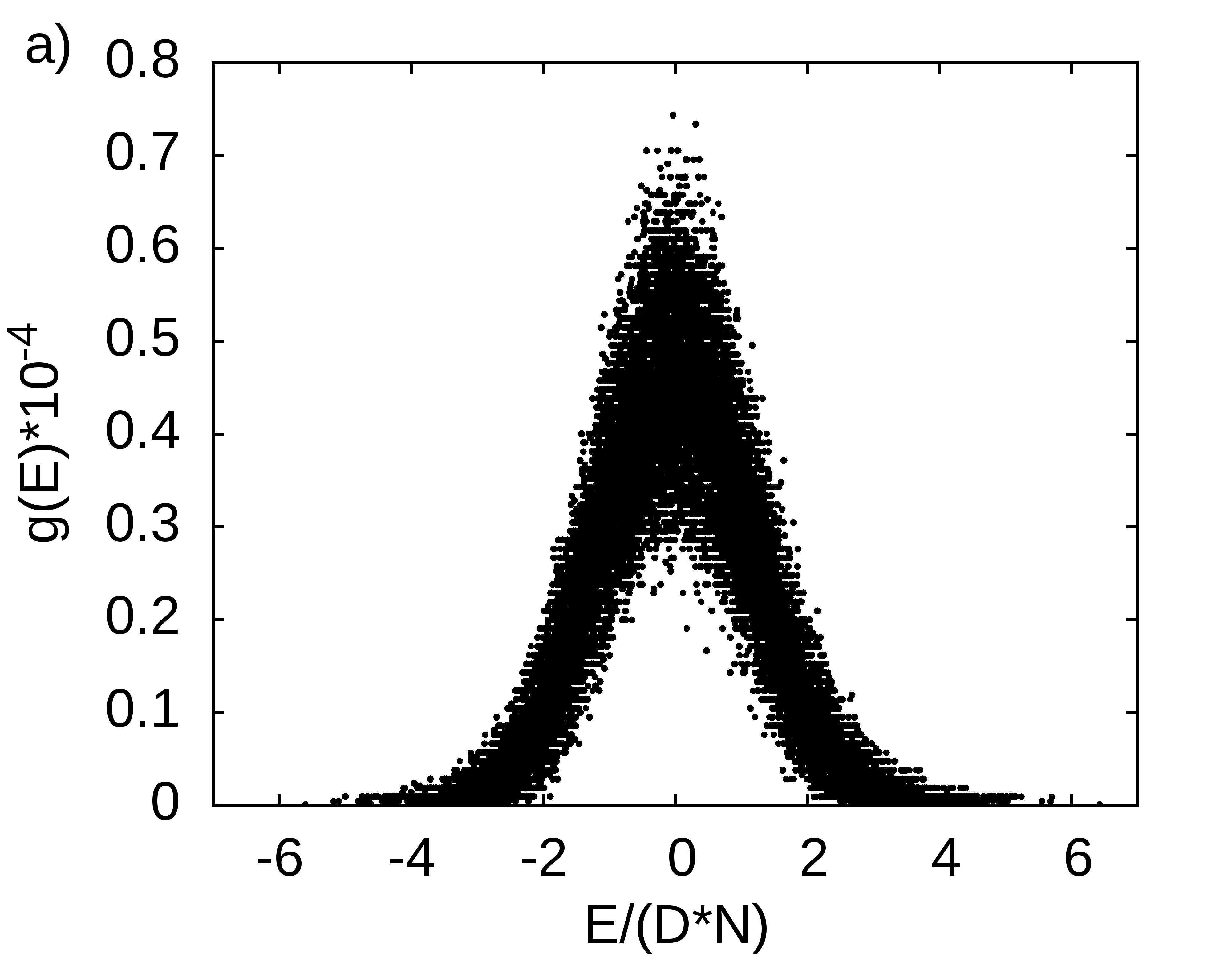}}\\
\end{minipage}
\hfill
\begin{minipage}[h]{0.49\linewidth}
\center{\includegraphics[width=1\linewidth]{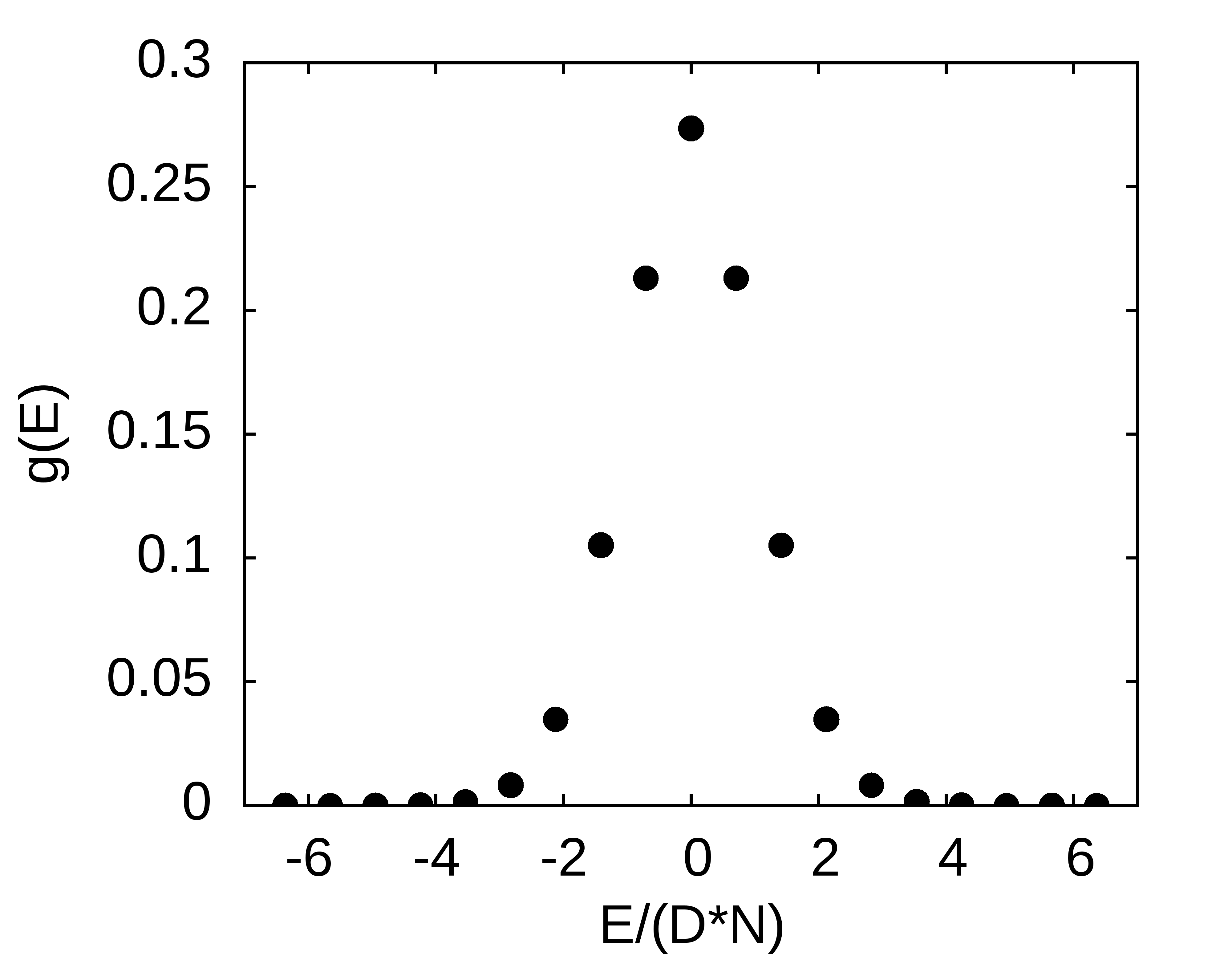}} \\
\end{minipage}

\vfill
\begin{minipage}[h]{0.49\linewidth}
\center{\includegraphics[width=1\linewidth]{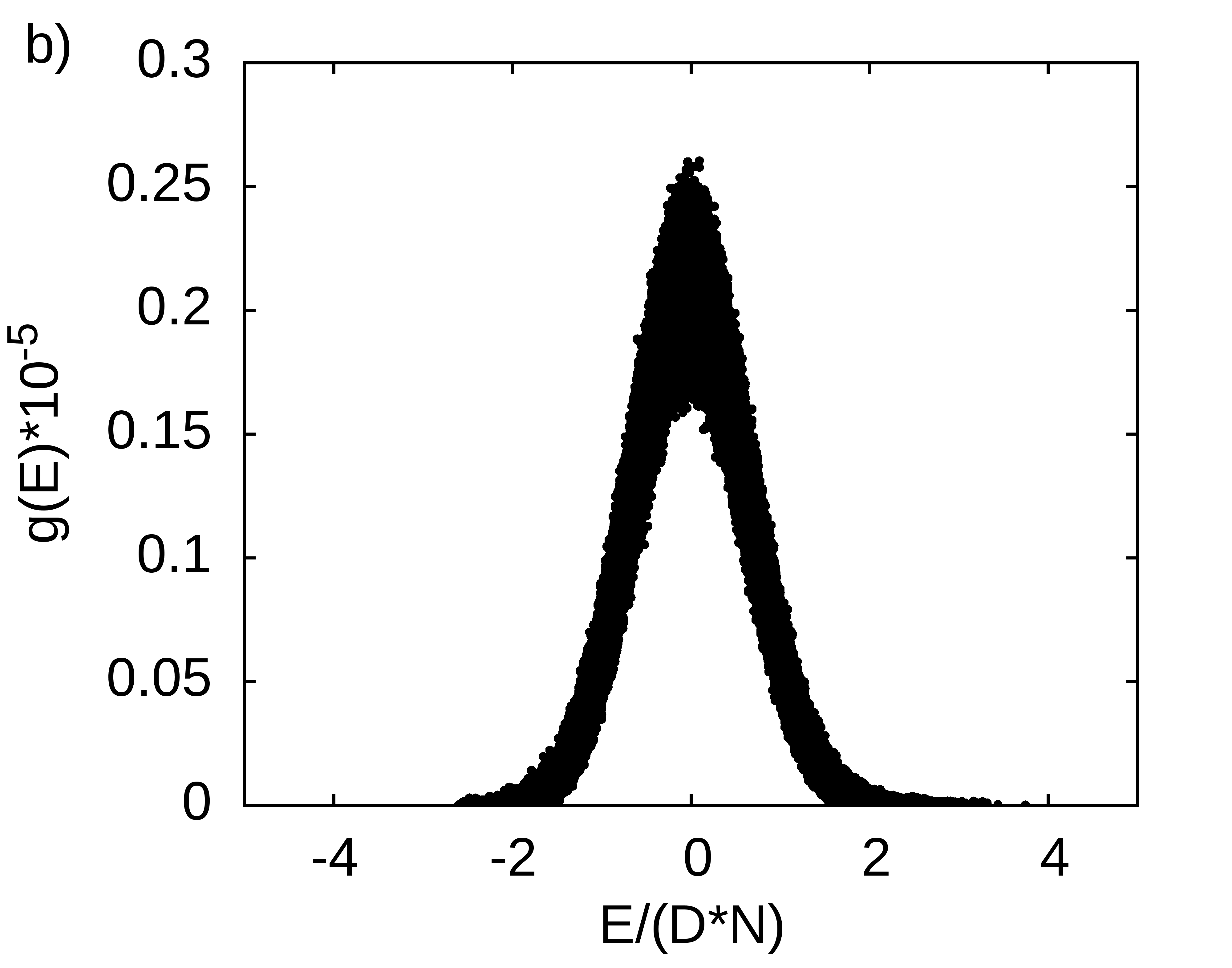}}\\
\end{minipage}
\hfill
\begin{minipage}[h]{0.49\linewidth}
\center{\includegraphics[width=1\linewidth]{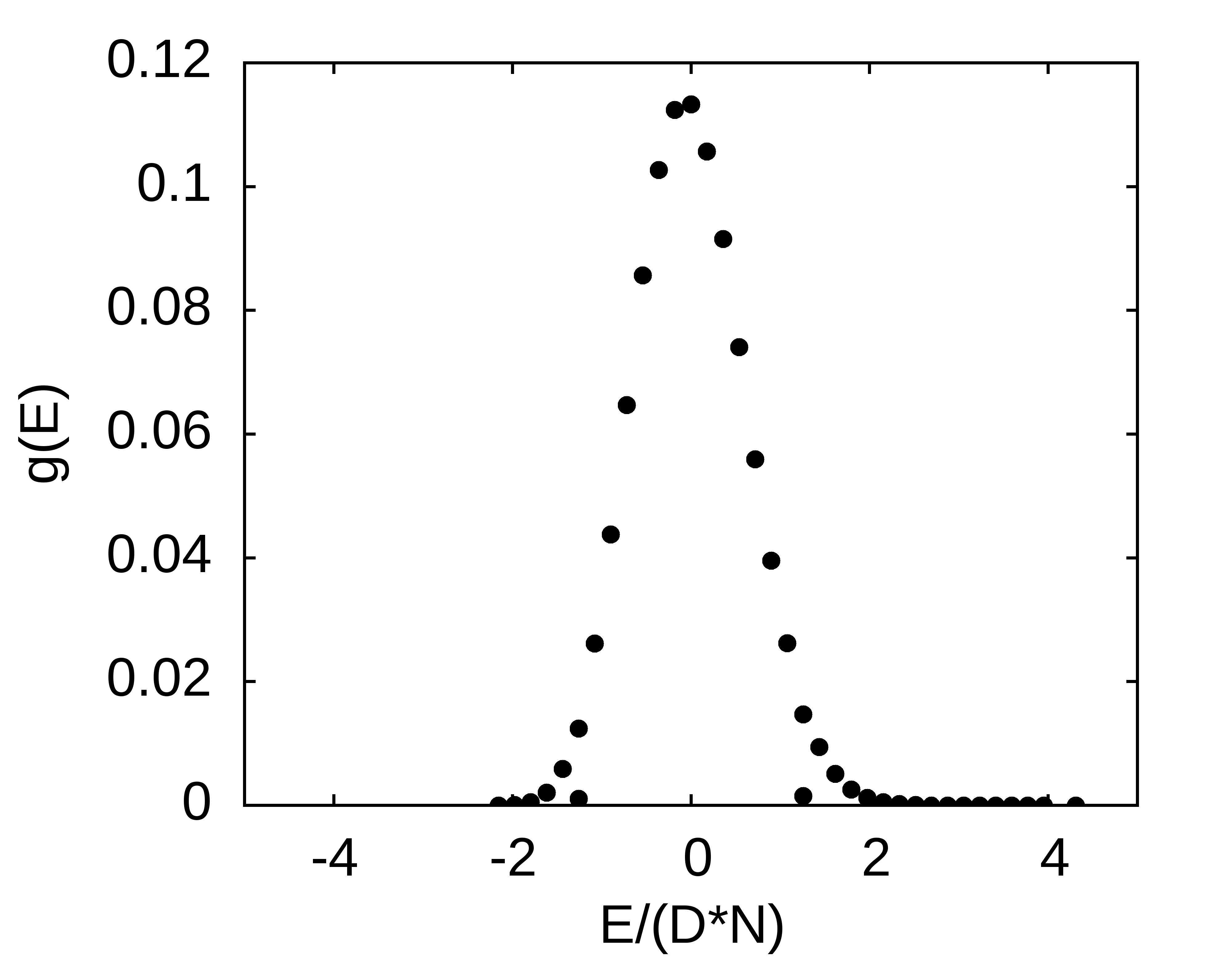}} \\
\end{minipage}

\begin{minipage}[h]{0.49\linewidth}
\center{\includegraphics[width=1\linewidth]{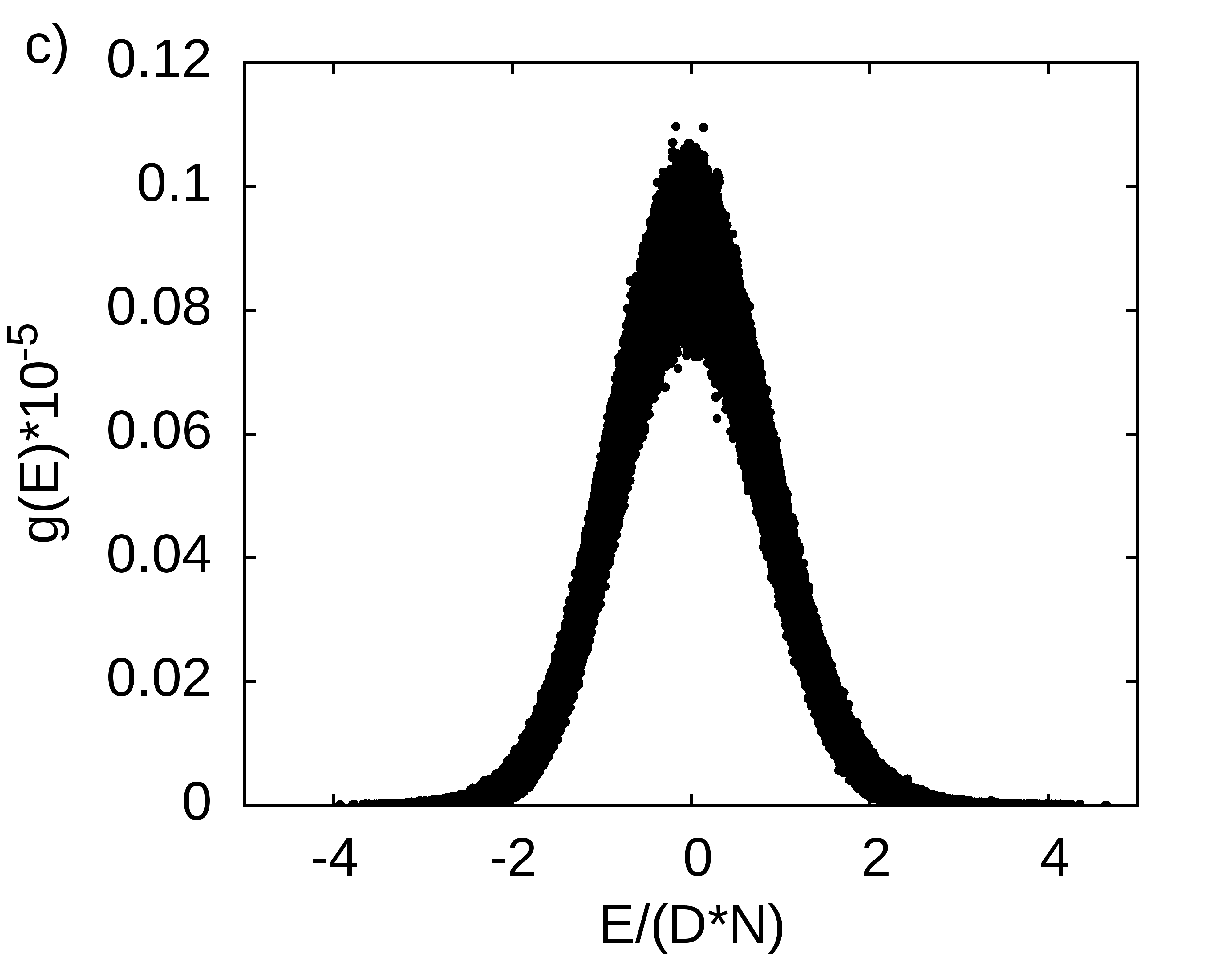}} \\
\end{minipage}
\hfill
\begin{minipage}[h]{0.49\linewidth}
\center{\includegraphics[width=1\linewidth]{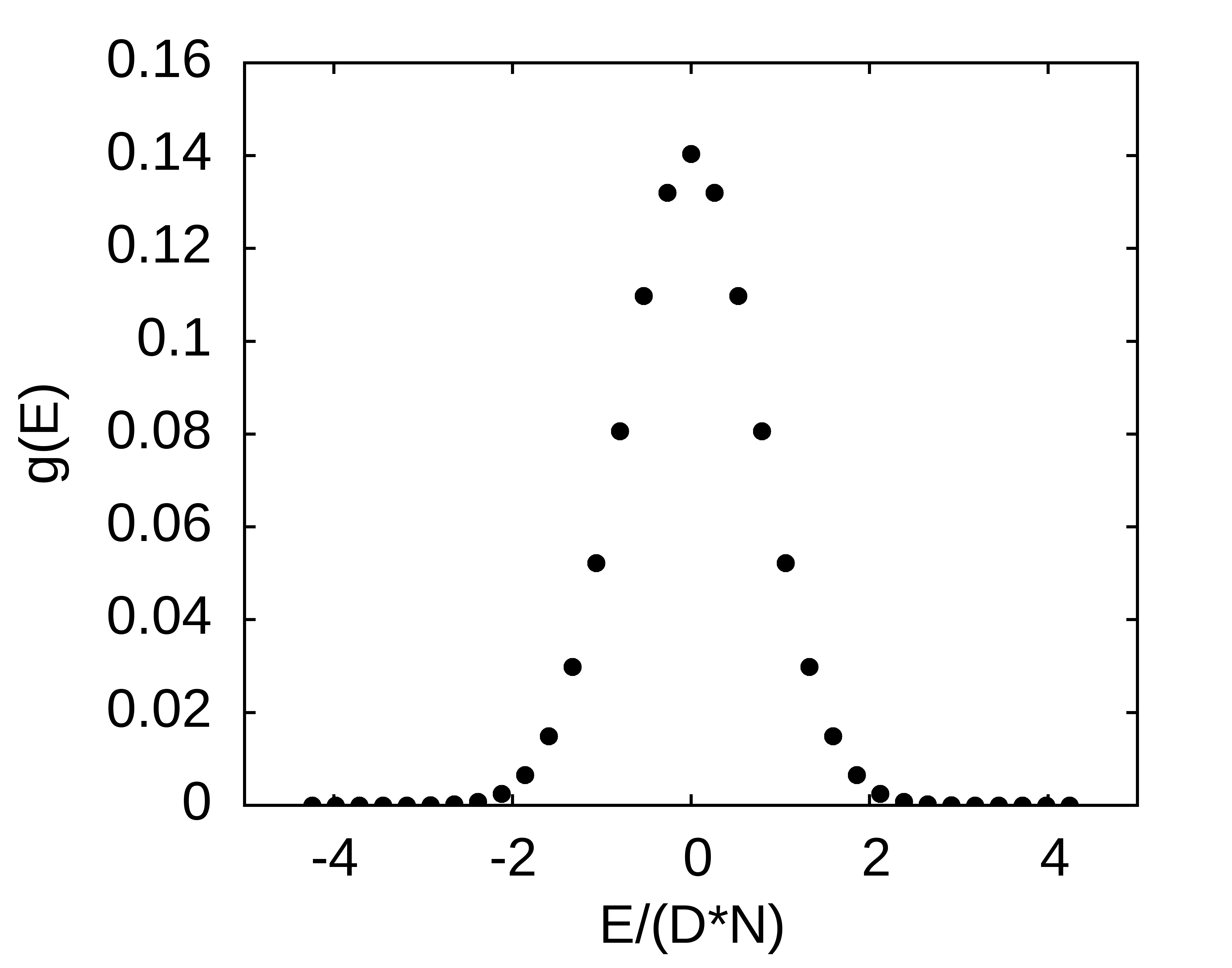}} \\
\end{minipage}

\caption{Density of states for long-range (left column) and short-range (right column). a) Square spin ice, 24 spins. b) Honeycomb lattice, 30 spins. c) Shakti lattice, 32 spins. Data obtained using the complete enumeration method over all space of states.}
\label{fig:dos_examples}
\end{figure}

\begin{figure*}
  \begin{minipage}[h]{0.32\linewidth}
  \center{\includegraphics[width=1\linewidth]{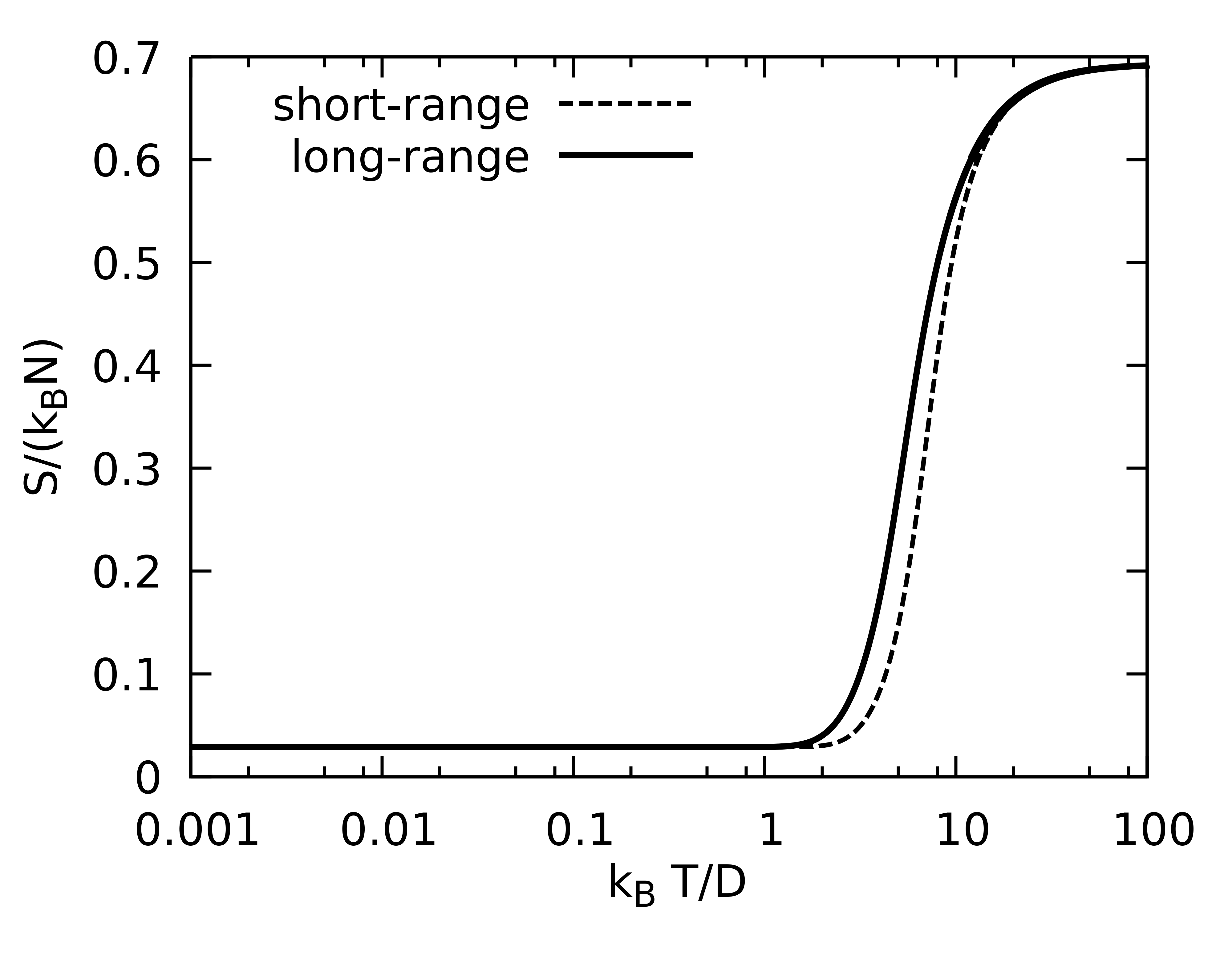}}
  \end{minipage}
  \hfill
  \begin{minipage}[h]{0.32\linewidth}
  \center{\includegraphics[width=1\linewidth]{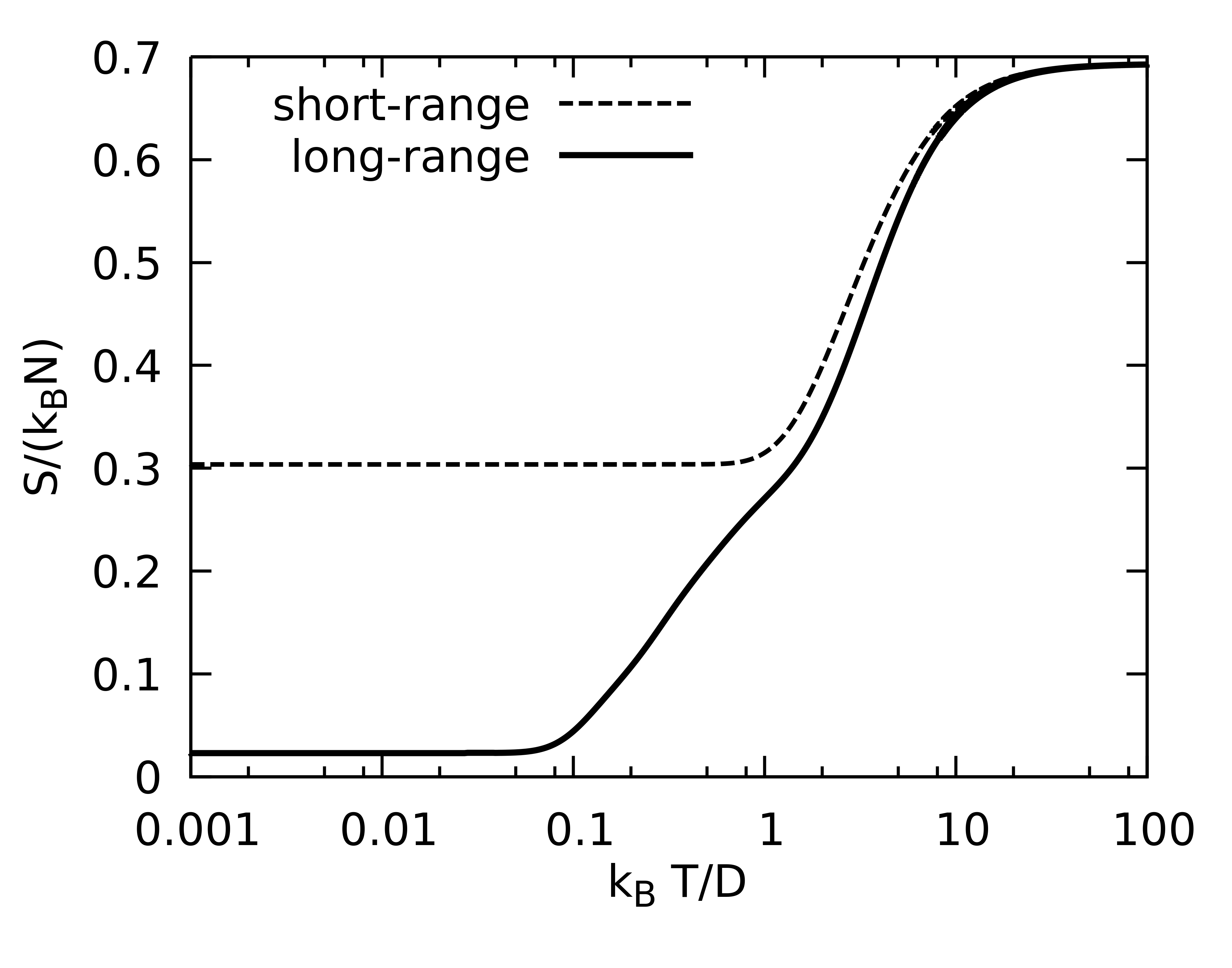}} 
  \end{minipage}
  \hfill
  \begin{minipage}[h]{0.32\linewidth}
  \center{\includegraphics[width=1\linewidth]{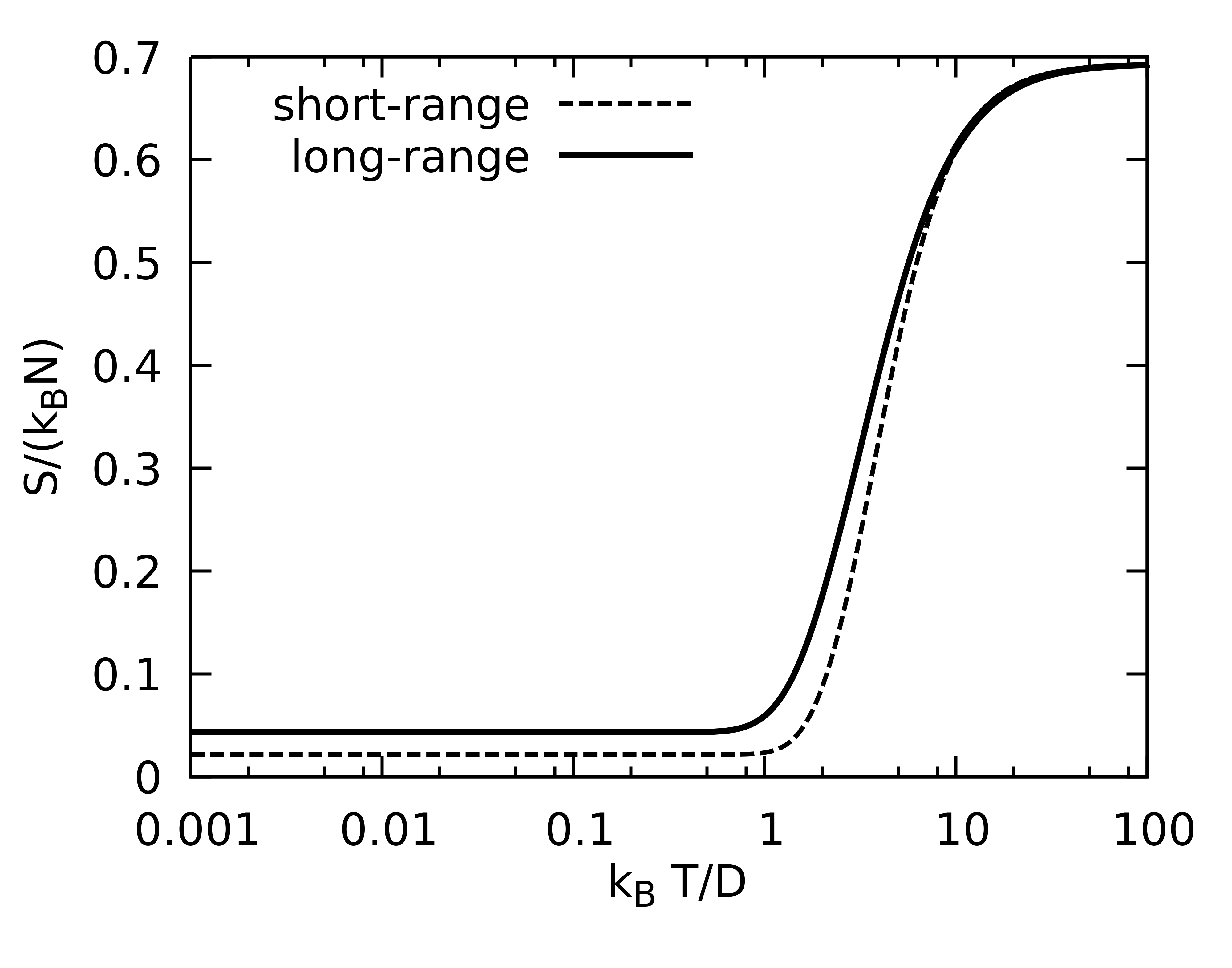}}
  \end{minipage} \\
  \vfill
  \begin{minipage}[h]{0.32\linewidth}
  \center{\includegraphics[width=1\linewidth]{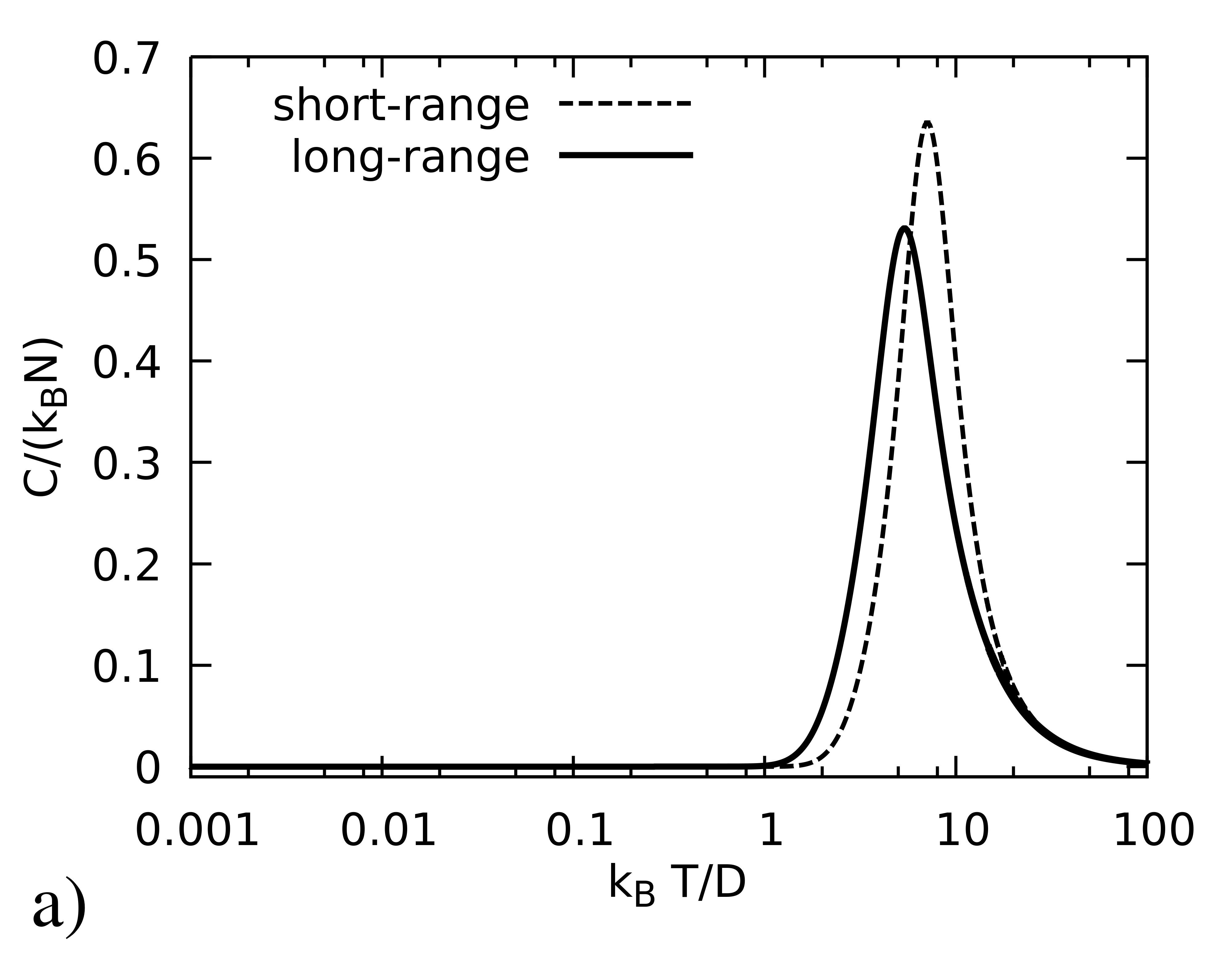}}
  \end{minipage}
  \hfill
  \begin{minipage}[h]{0.32\linewidth}
  \center{\includegraphics[width=1\linewidth]{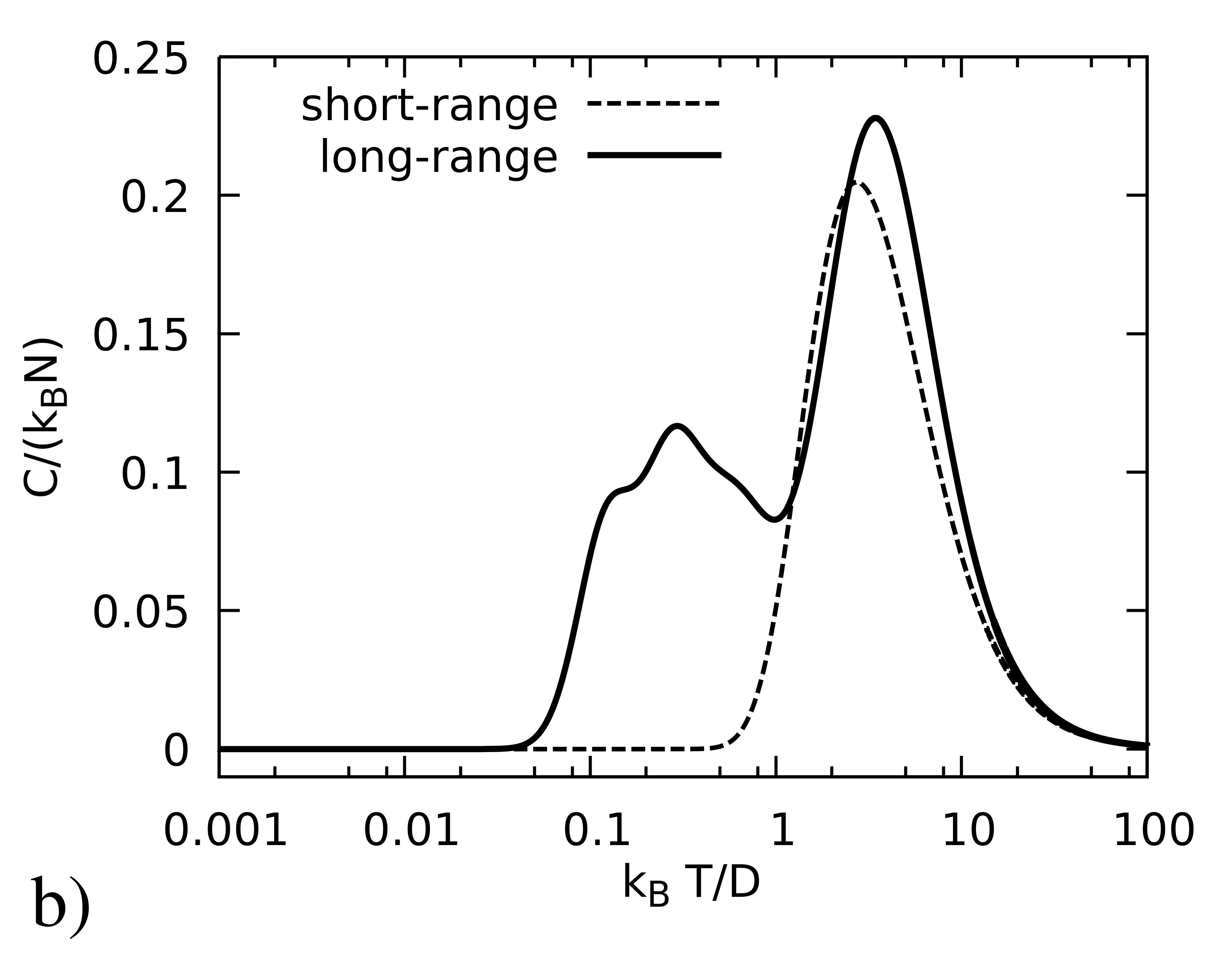}} 
  \end{minipage}
  \hfill
  \begin{minipage}[h]{0.32\linewidth}
  \center{\includegraphics[width=1\linewidth]{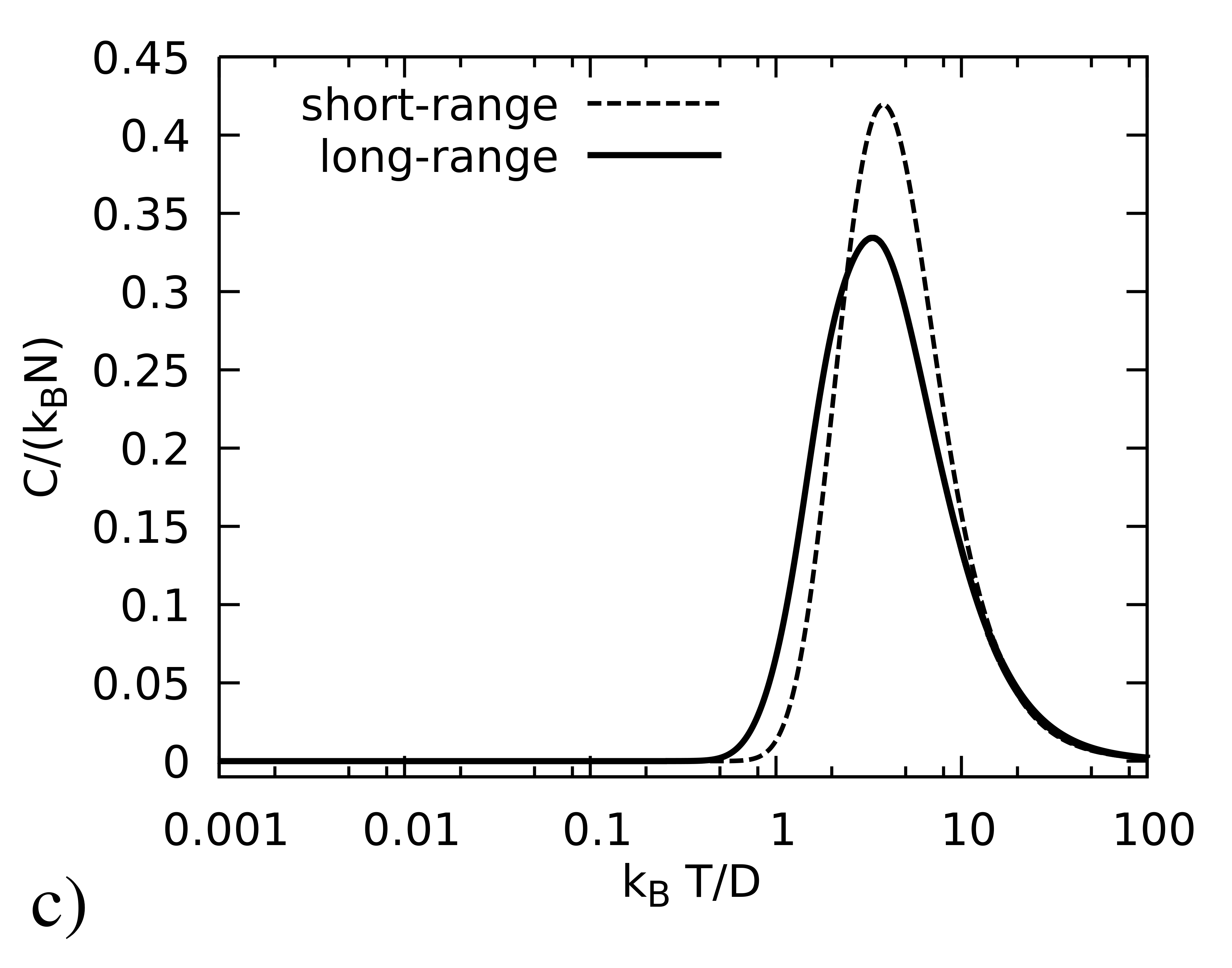}}
  \end{minipage} \\

  \caption{Entropy (top) and heat capacity (bottom). a) Square spin ice lattice, 24 particles. b) Honeycomb spin ice lattice, 30 particles. c) Shakti spin ice lattice, 32 particles.}
  \label{entropys}
\end{figure*}

During simulation of the long-range interaction between the spins of a system, certain restrictions are imposed on the total amount of numerical operations and data. 
The number of iterations required for the calculation of the energy of one configuration depends on the number of spins as $zN$, where $z$ is the number of nearest neighbours. 
It increases from $2N$ for $z=4$ at short-range interaction, to $N (N-1)$ for long-range interaction. 
The total number of possible configurations of the system is $2^N$. 
As a consequence, an exact calculation using the method of complete enumeration is only available for relatively small systems $N<40$. 
We use the complete enumeration to evaluate the density of states exactly (Figure \ref{fig:dos_examples}). 
The temperature dependencies of thermodynamic functions for ASSI, shakti-ASI and honeycomb-ASI are presented in Figure \ref{entropys}.
Large-size lattices of ASSI were investigated independently both by using the parallel Wang-Landau and Metropolis methods to guarantee convergence (Figures \ref{heat_312_24_long_short}, \ref{24_magnetization}, \ref{order_compare}). 
The correctness of the calculations is confirmed by comparing the simulation results to the exact solution for small $N$, and by comparing the results of independent numerical experiments using two MC methods for a relatively large number of particles.

We calculated the following thermodynamic values: the entropy $S(T)$, heat capacity $C(T)$, magnetic susceptibility in zero field $\chi_{\parallel,\perp}(T)|_{h \rightarrow 0}$ and the order parameter $\eta(T)$ as follows:
\begin{equation}
S(T)=-\frac{\partial F}{\partial T}=k_B \ln[Z]+ \frac{\langle E\rangle}{T},
\label{eq:entropy}
\end{equation}
\begin{equation}
C(T)= \frac{\langle E^2 \rangle - \langle E \rangle^2}{k_B^2 T^2},
\label{eq:heating}
\end{equation}
\begin{equation}
\chi_{\parallel,\perp}(T)|_{h \rightarrow 0} = \frac{\langle M^2 \rangle - \langle M \rangle^2}{k_B T},
\label{eq:suspectibility}
\end{equation}
\begin{equation}
\eta(T)=\frac{\langle N_{c} \rangle }{N} ,
\label{eq:order}
\label{ord_par_mc}
\end{equation}where $\langle N_{c} \rangle$ - the average number of spins forming the largest cluster, where all pair dipole interactions of nearest neighbours are minimised. 
The configuration of spins of the vertices in the largest cluster obeys the ice rule.

According to the rules of statistical mechanics, for a rigorous calculation of the thermodynamic parameters in equilibrium, it is necessary to obtain a partition function $ Z $. 
However, this is possible only if we have information about all the $ 2 ^ N $ states. It is clear that obtaining such information for large systems requires enormous resources, and therefore it is possible only for a small number of particles. 
Therefore, approximate statistic methods are used for relatively large systems.

There are various kinds of MC methods \cite{PhysRevB.90.220406,PhysRevE.57.1155,metropolis1953equation}. 
We use the Metropolis algorithm, described in \cite{landau2000guide}. The scheme using the parallel Wang-Landau algorithm (WL) is given in \cite{vogel2013generic}. 
The WL algorithm generates the histogram $g(E)$ during calculations. The thermodynamic averaged energy for a known distribution of density of states (DOS) $ g (E) $ is calculated as
\begin{equation}
\langle E \rangle = \frac{1}{Z} \sum_{i}g(E_i) E_i \exp \{ -\displaystyle\frac{E_i}{k_B T} \} ,
\label{avg_energy}
\end{equation}
where the normalisation factor, or the partition function, is:
\begin{equation}
Z=\sum_{j} g(E_j) \exp \{ -\frac{E_j}{k_B T} \} .
\end{equation}
It is required to perform a pre-assessment of energy boundaries $ E_ {min} $ (ground state) and $ E_ {max} $ for the building of $g(E)$.
To determine the degeneracy of states, we compare the current energy of configuration with all previous energies.
It is important to take into account the specifics of working with a computer representation of floating-point numbers during the WL-sampling.
The comparison of two identical $double$ numbers often works incorrectly. 
Each calculated energy will have a single, or at most, a double, degeneracy; the number of elements of the histogram will be very large.

To solve this problem, we divide the energy space between $ E_ {min} $ and $ E_ {max} $ on intervals. 
The minimum possible number of intervals was estimated in order to achieve good convergence.
The reduction of the intervals number may adversely affect the accuracy of the results; increasing it may rapidly enlarge the costs on calculations.
The width of the selected interval $ \triangle E $ should be large enough to show the energy landscape and restricted areas near the minimum or maximum adequately. 
At the same time, it should not be infinitely small, in order not to exceed the limit of computational accuracy and the amount of available computer memory.
We use $10^4$ intervals for square spin ice (Figure \ref{fig:dos_examples}a), $2*10^7$ intervals for honeycomb and shakti (Figures \ref{fig:dos_examples}b, \ref{fig:dos_examples}c). 
Therefore, the small number intervals for square spin ice is due to the small number of particles. 
The total amount of configurations is less than $2*10^7$. 
If we divide DOS into $2*10^7$ intervals, the landscape becomes flat (i.e. the similar energy values blur due to float computational errors). 

The results of our WL algorithm implementation are in good agreement with the exact solution. The accuracy for a relatively large number of particles is confirmed by the coincidence of heat capacities, susceptibility and order parameter, obtained by the WL and Metropolis algorithms independently.

\section{Exact solution for entropy and heat capacity}

Figures \ref{ASI_example}a, \ref{ASI_example}c and \ref{ASI_example}d show the ground states (GS) of ASSI, honeycomb-ASI and shakti-ASI lattices, respectively.
These GS are observed both in models with long-range and short-range interactions for the studied number of dipoles.
There are no frustrations for short-range interaction (for first coordination sphere) model ASSI (Figure  \ref{ASI_example}a).

For ASSI, each vertex obeys the ice rule \cite {Makarov2015126}. Exciting states, which also obey the ice rule (for example, all in rows are left and all in columns up), are not ground states. 
The ASSI has two ground states with opposite chirality, and thus a twofold degeneracy of the minimum energy is observed, regardless of the interaction radius. 
The existence of spins, the interaction energy of which is equal to zero (reversal without energy dissipation) was studied in \cite{Shevchenko201623, Shevchenko2016148}.

Figure \ref{fig:dos_examples} shows diagrams of the density of states obtained by the complete enumeration method over all possible configurations for the dipolar models. 
The right column of Figure \ref{fig:dos_examples} shows the model with truncated radius of dipole interaction, which does not exceed the radius of the first configuration sphere, comprising only nearest neighbours. 
The left column contains the density of states for models with long-range dipole interaction (i.e. full connected models).

In the density of states of the simple square spin ice, honeycomb and shakti lattices, we reveal the similarities in the shape of distribution between long-range and short-range models. 
However, the long-range interaction leads to a blurring of the density of states in all of the cases described here. We note a pronounced asymmetry of DOS for honeycomb \ref{fig:dos_examples}b in both ranges.

In all cases, we observe the Gauss-like distribution of DOS.
Naturally, the discrete spectrum of energy values is much poorer and there are significant band gaps. The forbidden energies in long-range are almost absent.

Frustration phenomena, to our understanding, represent the property of the equilibrium state, where the total energy cannot be completely minimised because of the existence of positive paired dipole interactions. 
Unsatisfied pair dipole-dipole interactions are present, even in the state with the minimum possible total energy.

Frustrations take place for shakti ice in the GS; the degeneration is also twofold. 
Frustrations are present for honeycomb, and degeneracy of the GS is a macroscopic.

Figure \ref{entropys} shows the entropy and heat capacity of square spin ice, honeycomb and shakti, obtained in the model of point dipoles by the complete enumeration method without any additional simplifications, scale limits, assumptions and admissions.
The number of spins is relatively small. 
This allow us to enumerate all possible states, and to calculate the partition function and thermodynamic potentials exactly. 

\begin{table}
  \begin{tabular}{lcclcc}
  \hline
      Lattice & 
      \begin{tabular}{@{}c@{}}spin\\count\end{tabular} & 
      Fig. & range & 
      \begin{tabular}{@{}c@{}}frust-\\rations\end{tabular} &
      \begin{tabular}{@{}c@{}}degenerations \\ of GS\end{tabular}\\
      \hline
    Square spin ice & 24 & \ref{entropys}a & short & no & 2 \\
    Square spin ice & 24 & \ref{entropys}a & long & yes & 2 \\
    Honeycomb & 30 & \ref{entropys}b & short & yes & $4000$ \\
    Honeycomb & 30 & \ref{entropys}b & long & yes & 2 \\
    Shakti & 32 & \ref{entropys}c & short & yes & 2 \\
    Shakti & 32 & \ref{entropys}c & long & yes & 4
  \end{tabular}
  \caption{Comparison of properties for lattices discussed above. The visualisation is shown on Figure \ref{ASI_example}. The degenerations of GS are taken from DOS. It should be noted that for any state in any lattice, it will always be the reverse state with the same energy. Thus, the degeneration level will also be a multiple of two.}
  \label{table1}
\end{table}

From Figure \ref{entropys} and Table \ref{table1}, it can be concluded that the long-range interaction may affect the presence of frustrations, residual entropy level, heat capacity peaks and shift in its temperature. 
The concrete effect depends on the lattice geometry and is weakly predictable.

The long-range dipole-dipole interaction not only blurs the density of states, and permits energy states that are prohibited in the short-range model, but also reduces the overall relative multiplicity of the degeneracy of GS. 
Surprisingly, the effect of long-range interaction, depending on the lattice topology, can lead to a shift in the critical temperature (i.e. the temperature of maximum heat capacity) both towards lower temperatures (square lattice and shakti) and higher temperatures (honeycomb). 
It should be noted that the form of the distribution function of the probability density of states is similar for all the considered lattices, and does not affect the appearance of features in the temperature dependence of the thermodynamic potentials for the honeycomb lattice.

\section{Phase transition in frustrated square spin ice}

\begin{figure}
\includegraphics[width=1\linewidth]{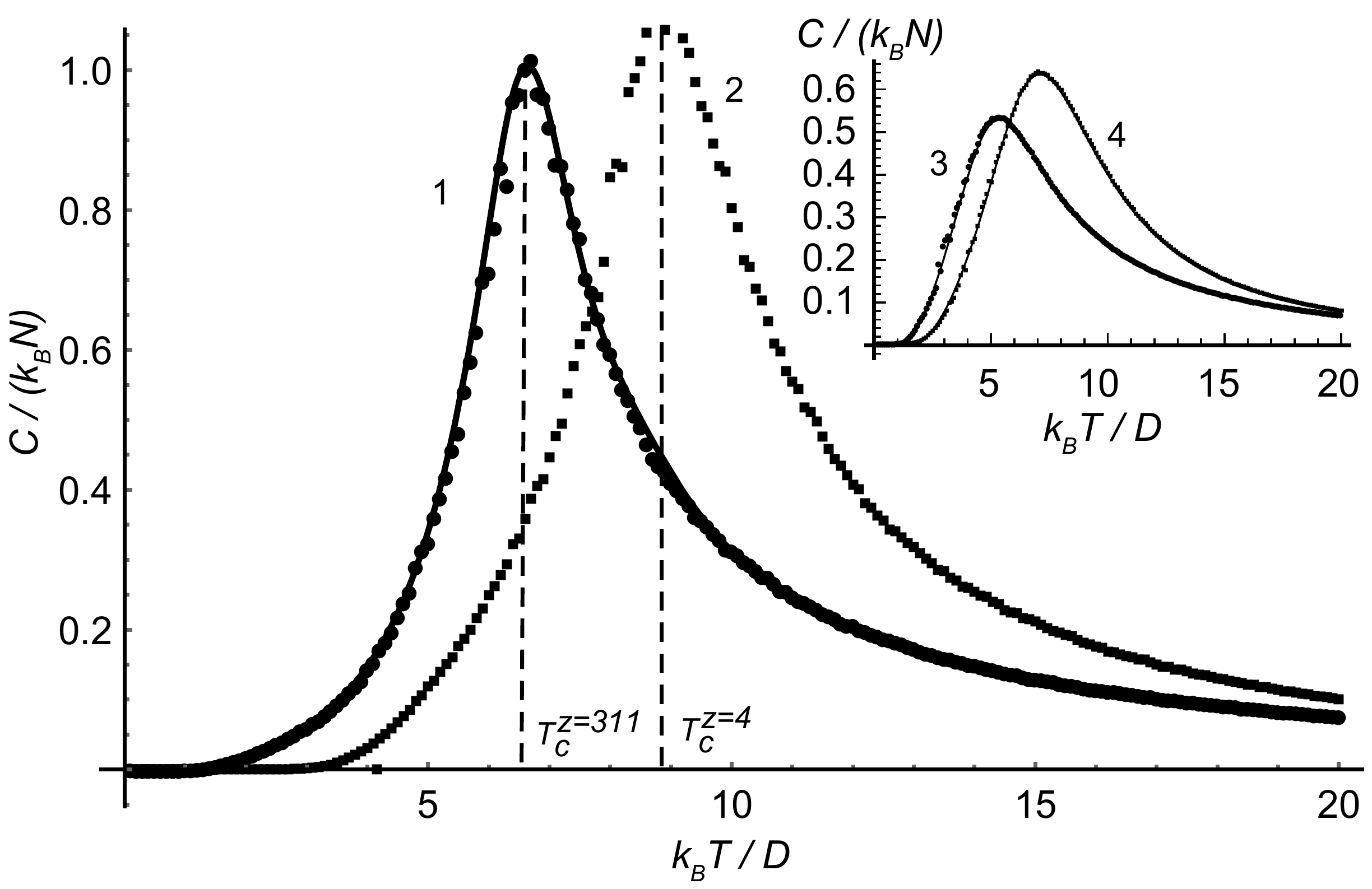}
\caption{Heat capacity of ASI for $N=312$. 
Solid line and point curves (1) - WL-sampling and Metropolis for the long-range model respectively, $z=311$. 
Point curve (2) - Metropolis for short-range interaction, $z=4$. 
The inset shows the rigorous solution of the heat capacity of ASI for $N=24$, combined with the data obtained by the Metropolis and the Wang-Landau methods for long- and short-range interaction; curves (3) and (4) respectively.}
\label{heat_312_24_long_short}
\end{figure}

The temperatures of the ASSI heat capacity peaks (obtained by the methods of WL and Metropolis) are different for the long-range and short-range models: $T_{c}^{z=311}<T_{c}^{z=4}$; Figure \ref{heat_312_24_long_short}.
The convergence of the results obtained by the Metropolis method with the exact solution and WL for $ N = 24 $, and the convergence of the curves WL and Metropolis for $N = 312$, confirm the accuracy of the results (inset of Figure \ref {heat_312_24_long_short}).

The height of the heat capacity peak for the ASSI model with long-range is below that of the short-range model. 
However, the dependence of the temperature behaviour of heat capacity on the particle number becomes less pronounced with an increase in the quantity of spins.

\begin{figure}
\includegraphics[width=1\linewidth]{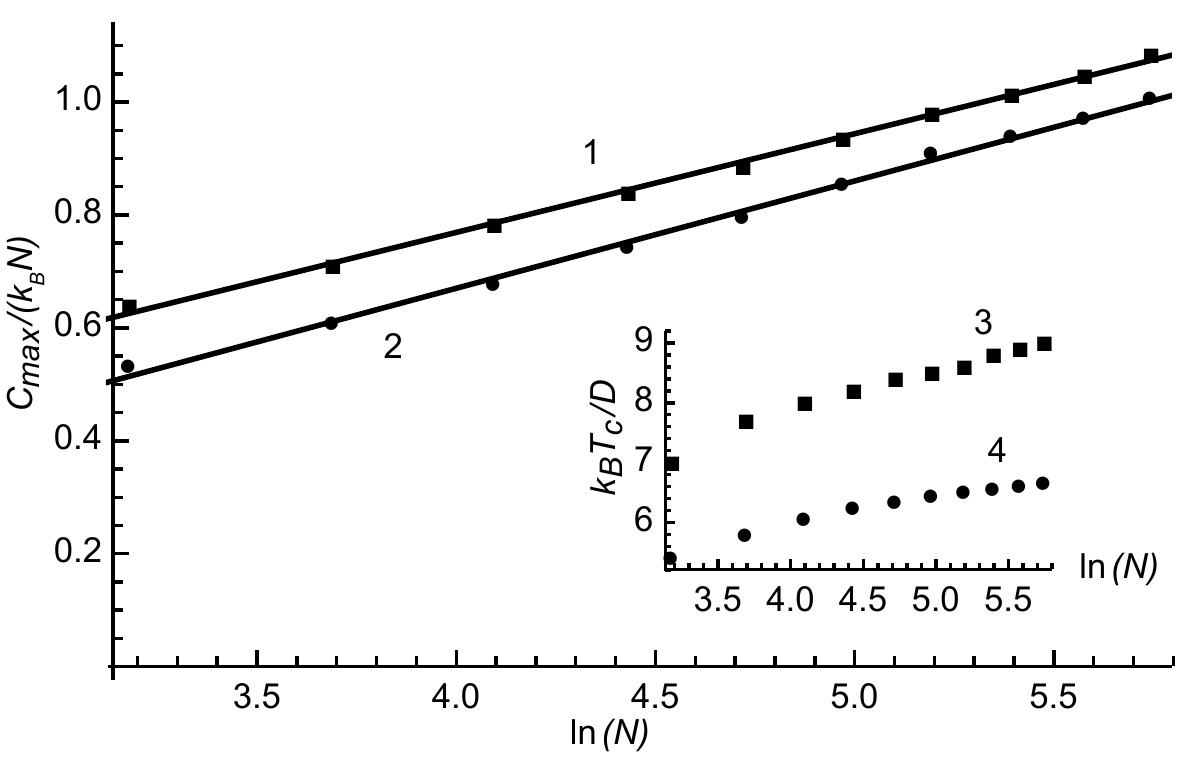}
\caption{Behaviour of heat capacity peak heights from the number of nanoislands. 
Insert shows the dependence of critical temperature $T_c$ on the number of nanoislands. 
Curves (1) and (3) were obtained by the Metropolis method for short-range interaction. 
Curves (2) and (4) were obtained by the WL method in the model for long-range interaction. 
}
\label{heating_peaks}
\end{figure}

Based on the results in Figure \ref{heating_peaks}, it may be assumed that the temperature behaviour of the heat capacity in the thermodynamic limit diverges at $T_c$ for both interaction lengths.

The possibility of a second-order phase transition from the paramagnetic phase to antiferromagnetic ''c''- type is confirmed by independent calculations in the $2D$ 16-vertex model and in the Bethe-Peierls approach \cite{levis2013thermal}.
The MC simulation with nearest-neighbours interaction converging on a single node $ z = 6 $ \cite{1367-2630-14-1-015008} also shows a sharp peak at a temperature approximately equal to 7.2 $ (D / k_B) $. At this temperature, the heat capacity value increases slowly logarithmically with the size of the system.

\begin{figure}
\includegraphics[width=1\linewidth]{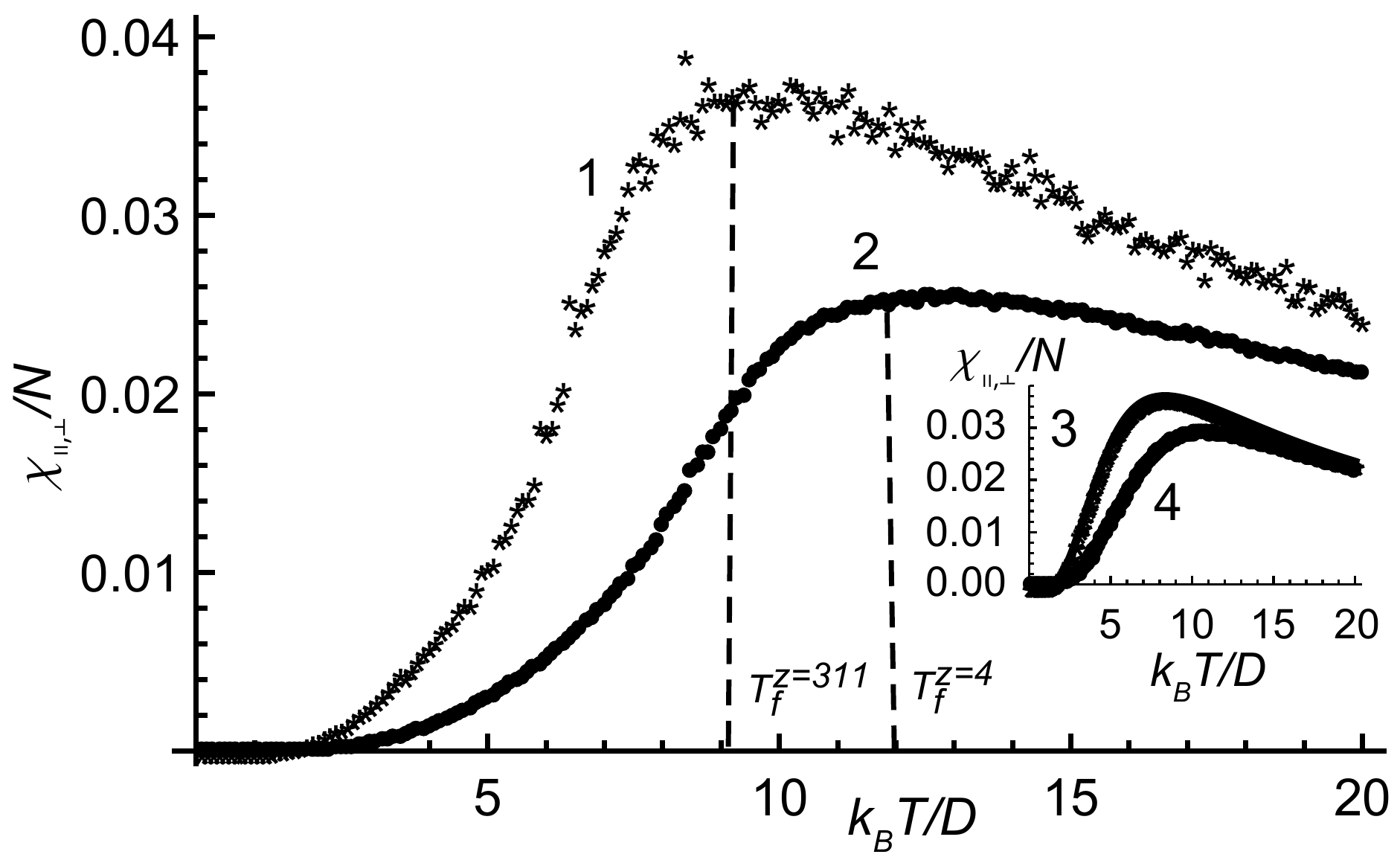}
\caption{The longitudinal (transverse) magnetic susceptibility in an infinitesimal external field for the $ N = 312 $ ASSI. 
Curve (1) - long-range interaction; curve (2) - short-range interaction, obtained by the Metropolis. 
Inset - rigorous solution for ASI with $N = 24$; curves (3) and (4) are calculated both numerically and using the Metropolis for two radii of interaction.}
\label{24_magnetization}
\end{figure}

The temperature behaviour of longitudinal (transverse) magnetic susceptibility $\chi_{\parallel,\perp}(T)|_{h \rightarrow 0}$ was obtained with Equation (\ref{eq:suspectibility}) using the Metropolis algorithm; Figure \ref{24_magnetization}. 
Curves (3) and (4) on the inset show the convergence with the results obtained by means of the complete enumeration method.
The calculation of $\chi_{\parallel,\perp}(T)|_{h \rightarrow 0}$ is fulfilled along one of two orthogonal sides of the square ASSI array.
Therefore, the main contribution to the susceptibility was produced by half the total number of moments that have the corresponding non-zero components of magnetisation. 
The other half of spins only have an indirect influence. 
$T_f$ is the temperature of the susceptibility peak.
For $\chi_{\parallel,\perp}(T)|_{h \rightarrow 0}$, and temperature of susceptibility peaks, $T_{f}^{z=311}<T_{f}^{z=4}$ the same as for heat capacity $T_{c}^{z=311}<T_{c}^{z=4}$. 
The susceptibility peaks for the long-range model were higher than for the short-range one.
It is necessary to note that for the studied finite-size samples of square ASSI, $T_c \neq T_f$.

In the square ASSI lattice, the mean value of magnetisation in the absence of an external magnetic field always equals zero at $T<T_c$ because of the features of the geometry. 
Therefore, this thermodynamic value cannot be used as an order parameter. 
Some suggestions about describing the order-disorder transition have been proposed. 
For example: the length of chains of co-directed spins, which are on the lattice bias (staggered magnetisation) \cite{levis2013thermal}; the relative number of flipped spins compared to the ground state configuration \cite{wysin2013dynamics}; all possible configurations of vertex, divided into four conditional types, and the relative part of each type at a given temperature, could be such a parameter \cite{wang2006artificial}.

\begin{figure}
\center{\includegraphics[width=1\linewidth]{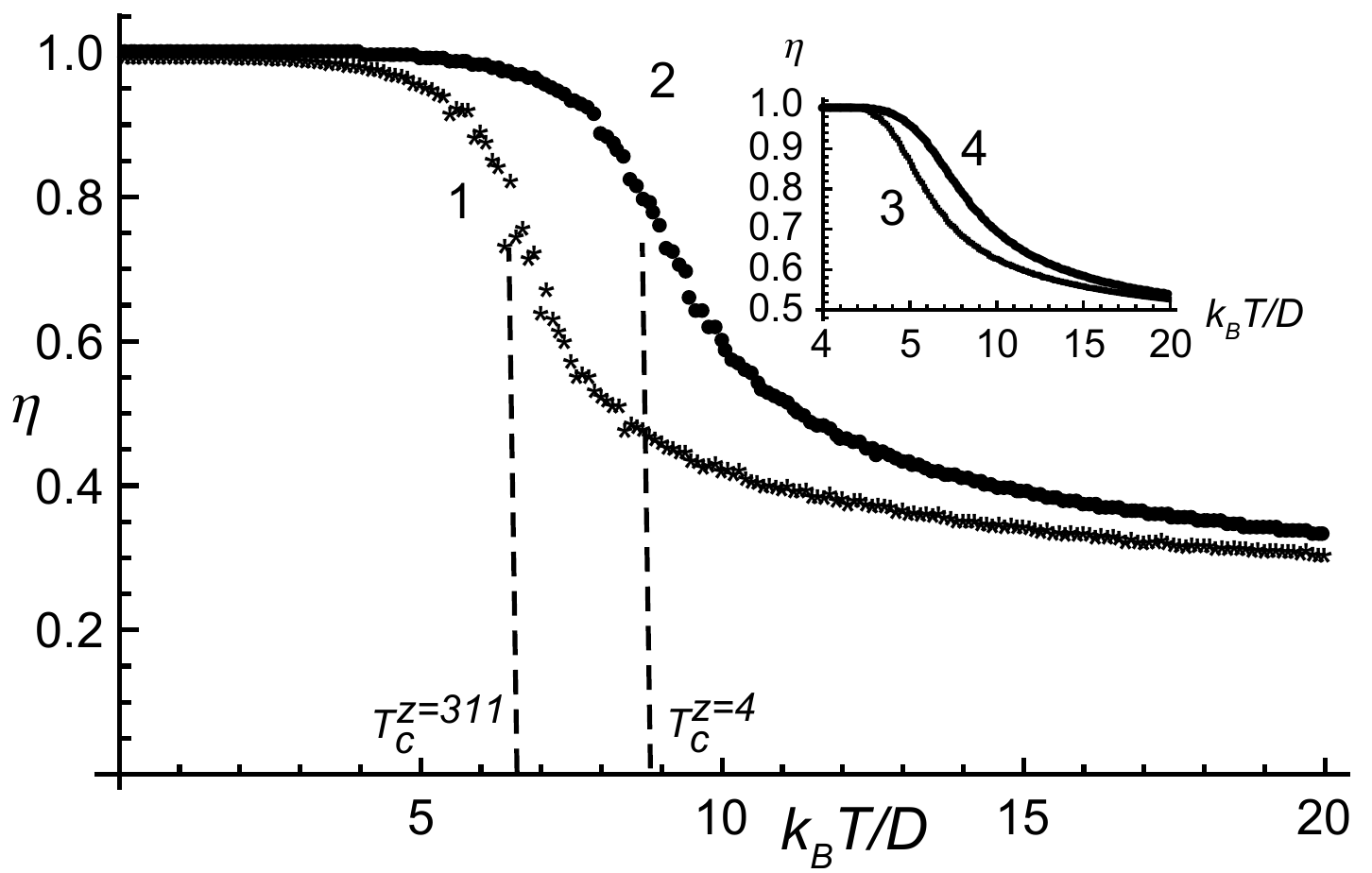}}
\caption{Comparison of order parameter $\eta$ of ASSI for $N=312$. At $z=311$, curve (1) and $z=4$, curve (2), respectively.
Data were obtained by means of the Metropolis algorithm. 
The inset shows the results of the exact solution for $N=24$, and the results of Metropolis sampling for the two studied radii of interaction (3) and (4).}
\label{order_compare}
\end{figure}

It is possible to suppose that the finite systems of ASSI spins at $T>T_f$ are in the paramagnetic regime. 
The nucleus of the new phase (i.e. the clusters of dipoles where the interaction energy is minimised) occurs in the temperature range $T_c < T < T_f$. 
Enabling the interaction at $T=T_f$ and its amplification at $T < T_f$ leads to a decrease in the response to an external magnetic field. 
On further lowering of the temperature, the probability of the appearance of such clusters increases. 
The clusters are combined and merged into one large cluster at $T=T_c$ (for infinite systems, it will be a percolation cluster and the largest one for finite systems).
At $T\rightarrow 0$, the probability that a randomly chosen spin belongs to largest cluster approaches 1.0.
The order parameter was thermodynamically averaged by Equation (\ref{ord_par_mc}). 
The ratio of the largest cluster size to the number of spins of the system is proposed as the order parameter $\eta$; Figure \ref{order_compare}. 

\section{Conclusions}
 We have investigated the long- and short- range dipole-dipole interaction effects on the frustrations, density of states, entropy and heat capacity in frames of square, honeycomb, shakti 2D spin ice lattices.
 
We observed the decreasing of critical temperatures in the behaviour of the thermodynamic parameters of the finite ASSI, comparing long- and short- range dipole-dipole interaction; the same situation for shakti-ASI. 
This can take place due to randomisation because of long-range dipole-dipole interaction, which prevents the emergence of order. 
The Curie temperature is the temperature of the heat capacity peak that corresponds to the temperature of maximum change of the average internal energy. For the honeycomb lattice, the temperature of the peak of heat capacity shifts to an area of high temperatures, while changing the short-range interaction into long-range. 
Moreover, two additional peaks arise in the temperature dependence of honeycomb heat capacity.

The reduction of the relative part of the low-energy states leads to a lowering of the Curie temperature in the long-range square spin ice model. 
As seen in Figure \ref{fig:dos_examples}, the long-range dipole-dipole correlations reduce the forbidden energy zones, increase the energy of the ground state, and significantly alter the distribution of DOS for the low-energy and excited states. Low-energy configurations bring the largest contribution to the Gibbs distribution if $T<T_f$. 
At $T \leq T_c$ the probability of a ground state configuration is suppressed, and equals $1.0$ for $T=0$. At $T_c<T<T_f$, the probabilities of low-energy configurations play the key role in the formation of the clusters. 
For this reason, the peak of magnetic susceptibility derivation is different to the heating peak.
At $T>T_f$ all configurations have the same probability, but that of a given value of $E$ is defined by DOS.

The heat capacity peak has a finite height due to the finite size effect. 
The increasing in size of the system leads to a slow logarithmic growth of the temperature of the heat capacity peak, both in the nearest-neighbour and the long-range interaction models (inset of Figure \ref{heating_peaks}).
It is interesting to note that the second-order phase transition is possible, even in systems with long-range dipole interaction. This conclusion is based on the data of Figure \ref{heating_peaks}, where the height of the heat capacity peak grows in both models.

The heat capacity peak temperature coincides with the temperature of the order parameter derivation peak (maximum changing speed of the largest cluster size). 
The condition for inclusion to the largest cluster was the minimisation of the interaction energy with the nearest neighbours placed in the first configuration sphere. 
The presence of the high-temperature tail in Figure \ref{ord_par_mc} is due to the finite size and free boundaries effects.

An interesting question for further investigation concerns reducing the impact of the border effect on the thermodynamic properties of systems with long-range interaction. 
In principle, it is easy to use periodical boundary conditions for long-range interaction in the way that each dipole has been in the same conditions over the entire system. 
However, the long-range interaction radius is limited by the size of the system in this approach. 
In addition, our clustering scheme (to include the nearest-neighbour spins that minimise the total energy) can be enhanced by taking into account the long-range interaction inside the cluster.

We assume that our proposed order parameter works not only for the models of the ASSI, but for magnetic structures with arbitrary geometry, and also for other condensed media, where the nature of the phase transition of the second order means the formation of clusters.

To study the collective behaviour of finite-size systems and their thermodynamics, it is important to understand the influence of the interaction range on the system characteristics. 
We conclude that the interaction length substantially affects the critical temperature, degeneration and frustrations. 
However, an open question is the so-called "effective length of the interaction", i.e. the distance at which the contribution of the long-range dipole-dipole interaction energy in the energy distribution will be insignificant.

\section*{Acknowledgements}
This work is financially supported by a grant from the President of the Russian Federation for young scientists and graduate students, in accordance the Program of Development Priority Direction ''Strategic information technologies, including the creation of supercomputers and software development'', grant \#SP-946.2015.5.

We would like to thank Professor Y. Okabe for his comments and for suggestions that led to substantial improvements.

\section*{References}
	
\bibliography{mybibfile}
\newpage

\end{document}